\documentclass[
    ,final            
    ,numberedheadings 
    ,cmfonts          
  ]
  {aipproc}

\layoutstyle{6x9}

\begin{document}

\title{Core-collapse supernova neutrinos and neutrino properties}


\author{J. Gava and C. Volpe}{
  address={Institut de Physique Nucl\'eaire Orsay, F-91406 Orsay cedex, FRANCE}\\
{E-mail: gava@ipno.in2p3.fr,volpe@ipno.in2p3.fr}
}
\begin{abstract}
Core-collapse supernovae are powerful neutrino sources. The observation of a 
future (extra-)galactic supernova explosion or of the relic supernova neutrinos
might provide important information on the supernova dynamics, on the supernova
formation rate and on neutrino properties. One might learn more about unknown
neutrino properties either from indirect effects in the supernova 
(e.g. on the explosion or on in the r-process) or from modifications of the neutrino time or energy
distributions in a detector on Earth. Here we will discuss in particular possible effects of CP violation
in the lepton sector. We will also mention the
interest of future neutrino-nucleus interaction measurements for the precise
knowledge of supernova neutrino detector response to electron neutrinos.

\end{abstract}

\maketitle


\section{Introduction}
Very massive stars produce violent gravitational explosions at the end of their life.
A huge amount of (anti-)neutrinos of all flavours is emitted during this phenomenon
that typically lasts a few seconds. The neutrino luminosity curve follows closely the
collapse, accretion phase and the cooling of the neutron star left. Stars with masses
larger than 20-25 solar masses can leave a black hole instead, with the neutrino
time distribution cut after a few hundreds milliseconds \cite{Beacom:2000qy}. 
Neutrinos emitted in Gamma-Ray-Bursts can also produce a characteristic neutrino signal \cite{McLaughlin:2006yy}.
Other interesting features of the supernova dynamics can be extracted from 
a core-collapse supernova neutrino signal like e.g. information on the shock-wave
\cite{Schirato:2002tg,Fogli:2004ff,Kneller:2007kg}. 

Core-collapse supernova neutrinos have been measured only once so
far, during the explosion of the SN1987A located in the Large Magellanic Cloud
at 50 kpc from us\footnote{This observation gave the Nobel Prize to M. Koshiba in 2002 with R. Davis for his pioneering experiment on solar neutrinos, and to R. Giacconi.}. The few events observed in the Kamiokande \cite{Hirata87}, IMB \cite{Bionta87} and
Baksan \cite{Alexeyev87} detectors have brought important confirmation on 
the theoretical 
expectations on the neutrino fluxes\footnote{Some features of the energy and
angular distributions remain to be understood.} and information on neutrino properties. 
Large scale observatories at present under study (LAGUNA Design Study financed within
the FP7) should reach the sensitivity to observe extra-galactic explosions and for 
the discovery of relic supernova neutrinos. 
Three technologies are under investigation: water \v{C}erenkov (MEMPHYS/UNO/Hyper-K), 
scintillators (LENA) and liquid argon (GLACIER) \cite{Autiero:2007zj}. About a hundred events are expected
in a detector such as Borexino, about 10000 in Super-Kamiokande \cite{Beacom:1998ya} and
10$^5$ in MEMPHYS for a typical supernova at 10 kpc \cite{Fogli:2004ff}. While explosions in our Galaxy are a rare phenomenon (typically 1-3 per century), the rate increases up to 1 per year at a distance as far
as 10 Mpc \cite{Ando:2005ka}. 

Relic supernova neutrinos (also called the diffuse supernova neutrino background) 
are neutrinos produced in past supernova explosions. The relic fluxes are not only sensitive
to the supernova dynamics and neutrino properties, but also on the star formation rate. 
The latter can be constrained through direct and indirect observations. At redshifts smaller
than 1 the rate is rather well known \cite{Ando:2004hc}; while
very recent works try to constrain at redshifts as high as 4 \cite{Yuksel:2008cu}\footnote{Note that if the supernova observatory has an energy threshold of about 10 (5) MeV, the flux is essentially due to z$<$1 (z$<$2).}.  The experimental present upper limits at 90 $\%$ C.L. are of 6.8 $\times$ 10$^3$ $\nu_e$ 
cm$^{-2}$s$^{-1}$  (25 MeV $< E_{\nu_e} <50$ MeV) and 1.2 $\bar{\nu}_{e}$ cm$^{-2}$s$^{-1}$  
($E_{\bar{\nu}_e} > 19.3$ MeV) come from the LSD \cite{Aglietta:1992yk} and
the Super-Kamiokande detectors \cite{Malek:2002ns} respectively. Present theoretical predictions show that the relic fluxes are very close
to the Super-Kamiokande limits, so that they might be discovered with large size detectors  \cite{Ando:2004hc}.  
In \cite{Volpe:2007qx} it has been proposed to improve our present relic electron neutrino limit by
using the few events associated to electron scattering on oxygen in MEMPHYS
or on carbon in LENA. On the other hand several hundred events might be
measured in GLACIER thanks to the large neutrino-argon
cross section \cite{Volpe:2007qx}, as first pointed out in \cite{Cocco:2004ac}.  

Impressive progress has been made in neutrino physics in the last
ten years, after the discovery of the phenomenon of neutrino oscillations performed by the Super-Kamiokande experiment \cite{Fukuda:1998mi}. However crucial questions remain open, among which the 
value of the third neutrino mixing angle, of the CP violating phase(s), of the (Majorana versus Dirac) nature of neutrinos and of the absolute
mass scale. The study of neutrinos from the early Universe or from stars  brings important elements to progress on some of the open issues. In particular, unknown neutrino properties might have an indirect impact
on the supernova environment such as e.g. on the nucleosynthesis of the heavy elements during the r-process
or leave an imprint on the time or energy supernova neutrino signal in a detector on Earth.
For example 
in \cite{Engel:2002hg} we have shown that in
lead-based detectors one can exploit the measurement of
the electron neutrino 
average energy through one and two neutron emission
to extract information on the third neutrino 
mixing angle. A lead-based detector (the HALO project) is planned for construction 
at SNOLAB \cite{halo}.
Here we will discuss possible CP violating effects in the supernova context \cite{Balantekin:2007es}. 

In order to use supernova neutrinos to extract information on neutrino
properties, we need to progress on the uncertainties that still affect the 
supernova neutrino predictions.
In fact, the theoretical calculations of the neutrino fluxes are affected 
by uncertainties at
the neutrinosphere, due to different supernova modelling.
These uncertainties will narrow down once supernova explosions will be successfully obtained 
for different stellar masses and the relevant microscopic processes included in the neutrino
diffusion. Obviously, the number of unknowns can also reduce if  
planned neutrino experiments progress on some of the neutrino properties 
like e.g. 
if Double-Chooz \cite{Ardellier:2006mn} or Daya-Bay \cite{Guo:2007ug} measure the third neutrino mixing angle. 
On the other hand, recent important developments have also shed a new light on our understanding of neutrino propagation
in dense media in presence of the neutrino-neutrino interaction (see e.g. \cite{Pantaleone:1992eq,Samuel:1993uw,Sigl:1992fn,Pastor:2001iu,Fuller:2005ae,Duan:2006an,Raffelt:2007xt,Fogli:2007bk,Hannestad:2006nj}. In fact, significant differences arise in some cases
compared to the standard Mykheyev-Smirnov-Wolfenstein
effect \cite{Wolfenstein:1977ue,Mikheev:1986wj}. In the detection process,
one also need to take into account the knowledge of the detector response. If we aim at disentangling the different pieces of information,
it is obviously essential that we extract as much information as possible from future observations. 
While neutral current events exploiting the scattering on electrons or on protons \cite{Beacom:2002hs} will measure the total neutrino fluxes, sensitive to the normalisation factors; charged-current events will be flavour sensitive to both electron neutrino and anti-neutrinos. 
If electron anti-neutrinos can be detected through their interaction on
protons, that is known
theoretically at all orders, the measurement of electron neutrinos require interaction on nuclei that 
is affected by theoretical uncertainties. Here we will briefly discuss the need for future measurements with intense low energy neutrino beams.

\section{Possible CP violation effects in supernovae}
The possible existence of CP violation in the lepton sector is one of the crucial open issue
in neutrino physics that can have tremendous impact on other domains as well. If the value of the third neutrino mixing angle turns out to be very
small, only "third generation" accelerator neutrino experiments, namely super-beams, beta-beams and neutrino factories \cite{Volpe:2006in}, will reach the necessary sensitivity to explore this question.
\begin{figure}
  \includegraphics[height=.25\textheight]{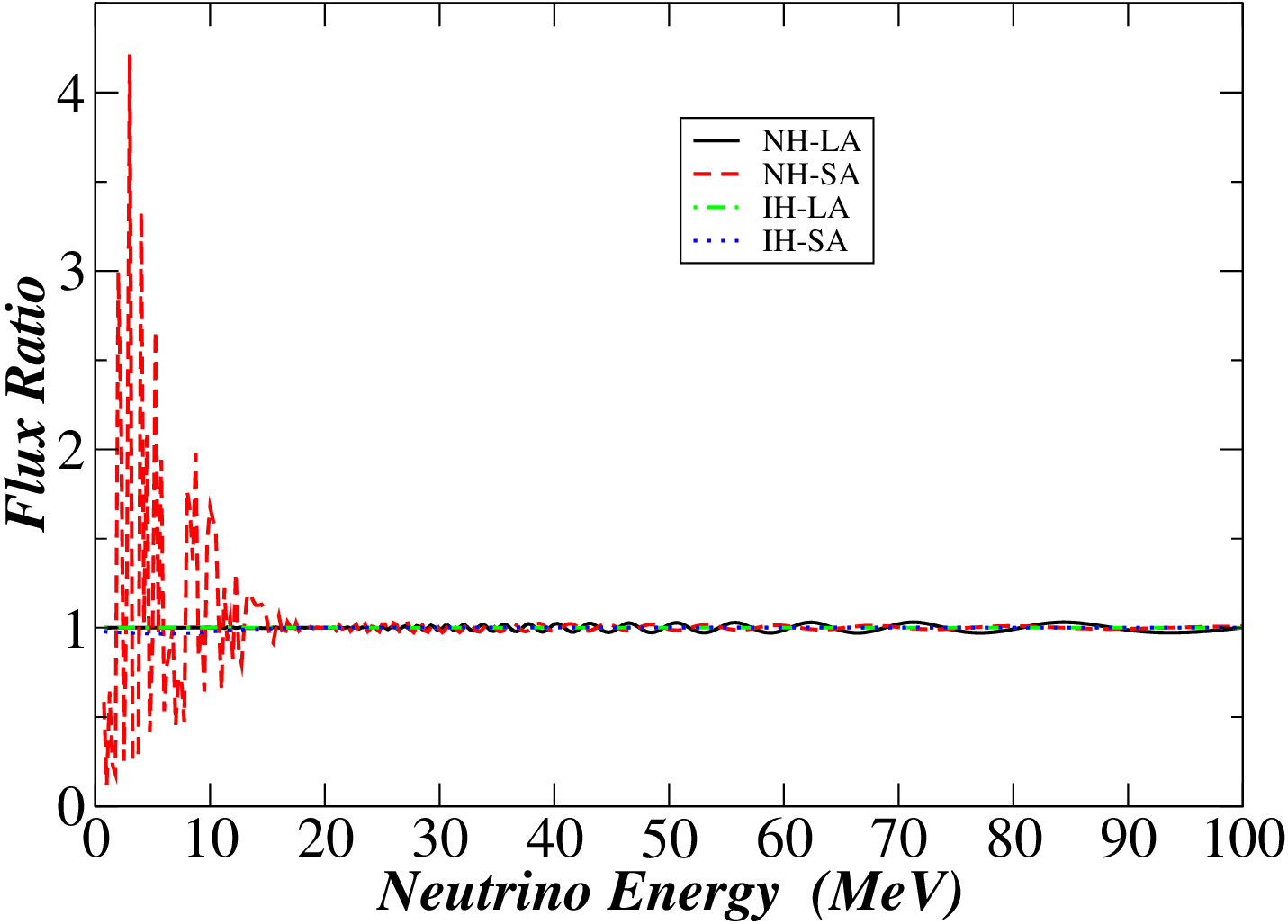}
  \caption{Muon neutrinos flux ratios for a CP violating phase $\delta=180^{\circ}$ and $\delta=0^{\circ}$ at 1000 km
from the neutrinosphere. The different curves correspond to the cases of
normal (NH) or inverted (IH) hierarchy and large (sin$^2 2\theta_{13}=0.19$,
LA) or small (sin$^2 2\theta_{13}=3 \times 10^{-4}$,SA) third neutrino mixing angle \cite{Balantekin:2007es}. \label{fig:nutau}}
\vspace{.2cm}
\end{figure}
It is important to search for other strategies to learn about the value of the CP violating phase, like for example
with neutrino astronomy \cite{Winter:2006ce,Akhmedov:2002zj}. 
In a recent work \cite{Balantekin:2007es} we have investigated both
analytically and numerically possible CP violating effects in the supernova
environment. The neutrino fluxes within the supernova are obviously
\begin{equation}\label{e:flux}
{\phi}_{\nu_i}(\delta) =  L_{\nu_i}P(\nu_i \rightarrow \nu_i) + 
L_{\nu_j}P(\nu_j \rightarrow \nu_i)+L_{\nu_k}P(\nu_k \rightarrow \nu_i)
\end{equation}
with the luminosities at the neutrinosphere $L_{\nu_i}, (\nu_i=\nu_e,\nu_{\mu}, \nu_{\tau})$ 
given by either Fermi-Dirac or power-law distributions, 
\begin{figure}
  \includegraphics[height=.25\textheight]{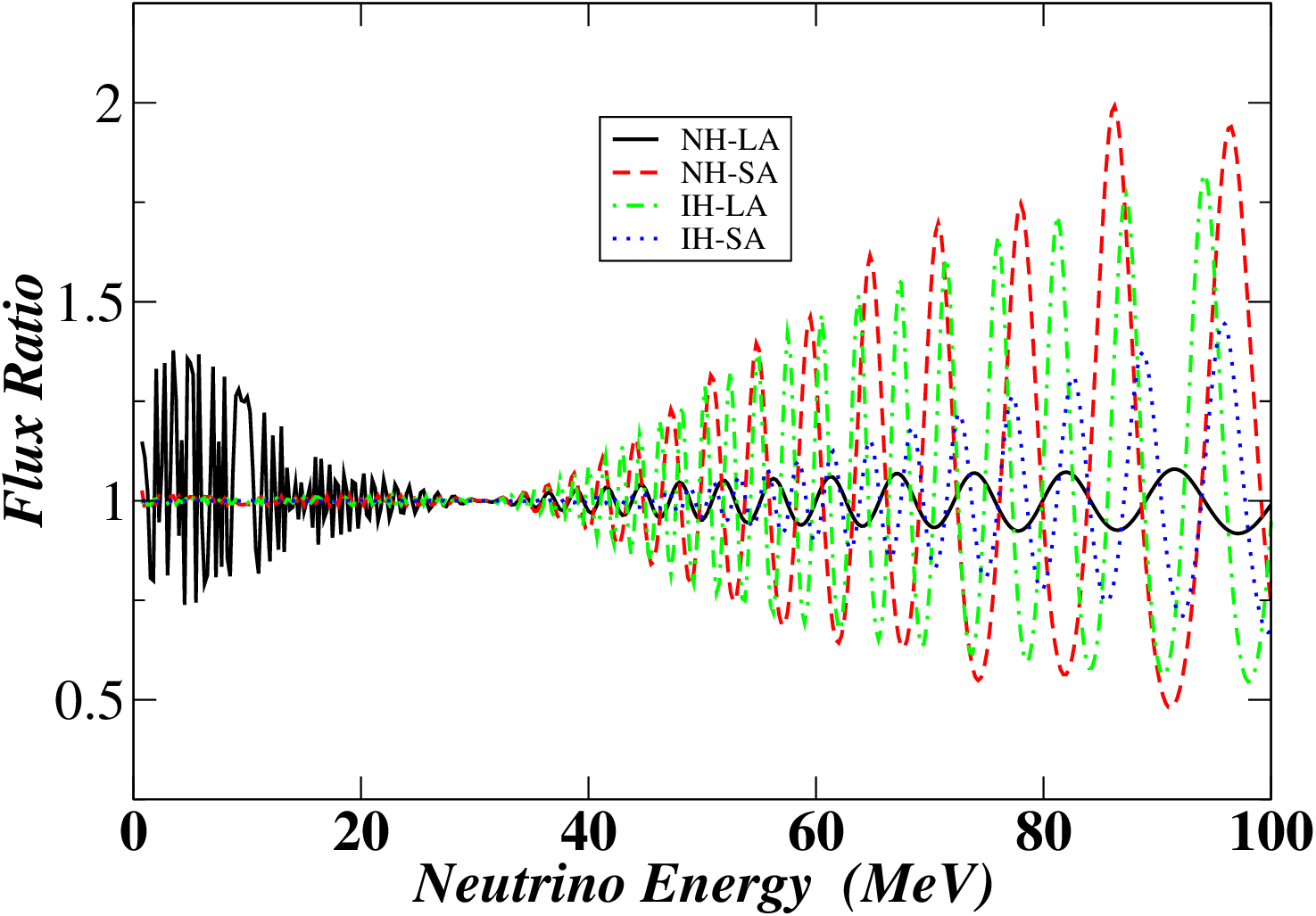}
  \caption{Same as Figure \ref{fig:nutau} but for electron neutrino fluxes. In this case $\nu_{\mu}$ and 
$\nu_{\tau}$ neutrino fluxes are taken different at the neutrinosphere by assuming that their temperatures differ by 1 MeV as an example \cite{Balantekin:2007es}. \label{fig:nue}}
\vspace{.2cm}
\end{figure}

\begin{figure}[t]
  \includegraphics[height=.25\textheight]{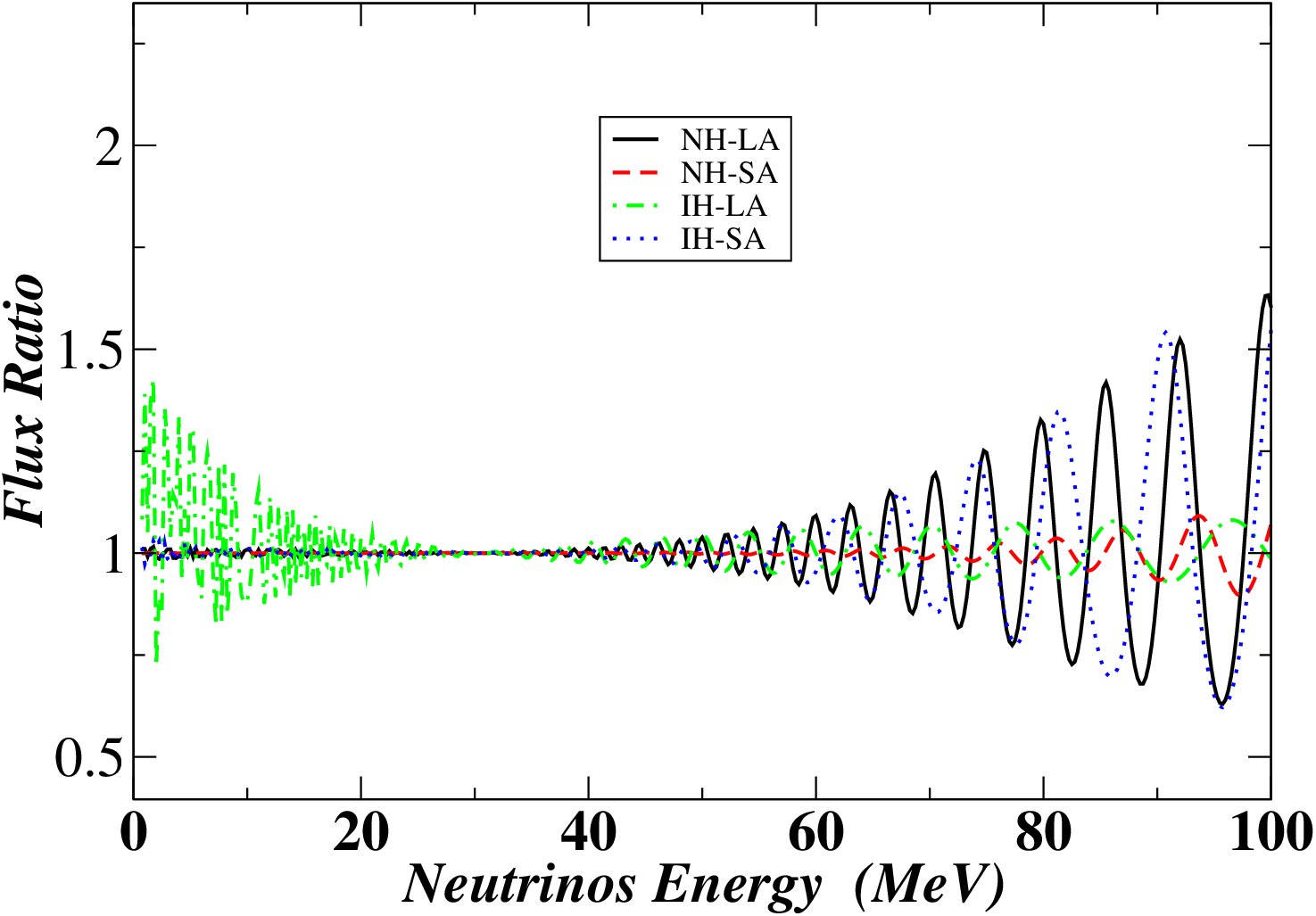}
  \caption{Same as Figure \ref{fig:nue} but for the electron anti-neutrino fluxes \cite{Balantekin:2007es}. \label{fig:anue}}
\vspace{.2cm}
\end{figure}
Analytically one can show that the following equation holds \cite{Balantekin:2007es}:
\begin{equation}
\label{sq2}
P (\nu_{\mu} \rightarrow \nu_e, \delta \neq 0) + 
P (\nu_{\tau} \rightarrow \nu_e, \delta \neq 0) =
P (\nu_{\mu} \rightarrow \nu_e, \delta =0) + 
P (\nu_{\tau} \rightarrow \nu_e, \delta =0) .
\end{equation} 
If $\nu_{\mu}$ and $\nu_{\tau}$ have equal luminosity at the neutrinosphere (true at tree level), 
Eqs.(\ref{e:flux}) and (\ref{sq2}) imply that there are no CP violation effects 
on the electron (anti-)neutrino fluxes for any density profile, as in vacuum. 
This conclusion is in agreement with the findings in \cite{Akhmedov:2002zj}. 
However higher order corrections to neutrino scattering on matter and/or physics  
beyond the Standard Model such as flavour changing interactions can
differentiate $L_{\nu_{\mu}}$ and $L_{\nu_{\tau}}$. 
In this case
the electron (anti-)neutrino fluxes do depend on the CP violating phase
Eqs.(\ref{e:flux})-(\ref{sq2}). 
 
We have performed for the first time numerical calculations of the neutrino propagation in a dense
environment with $\delta$ (for details see \cite{Balantekin:2007es}). 
Figure \ref{fig:nutau} presents the effects on the $\nu_{\tau}$ fluxes as an example.
As one can see the presence of the phase modifies them significantly, even
though the effects on the muon and tau neutrino fluxes
unfortunately cancel out once one consider neutral current interactions since 
all neutrino fluxes
add.  Figure \ref{fig:nue} and \ref{fig:anue} show the effects on the
electron neutrino and anti-neutrino fluxes
respectively once we assume that $L_{\nu_{\mu}} \neq L_{\nu_{\tau}}$. 
Such effects are interesting since the electron (anti-)neutrinos play a specific role both for the nucleosynthesis of heavy elements during the r-process and for a possible signal in a detector 
where electron (anti-)neutrinos can be measured through charged-current events.

\begin{figure}[t]
\includegraphics[height=.25\textheight]{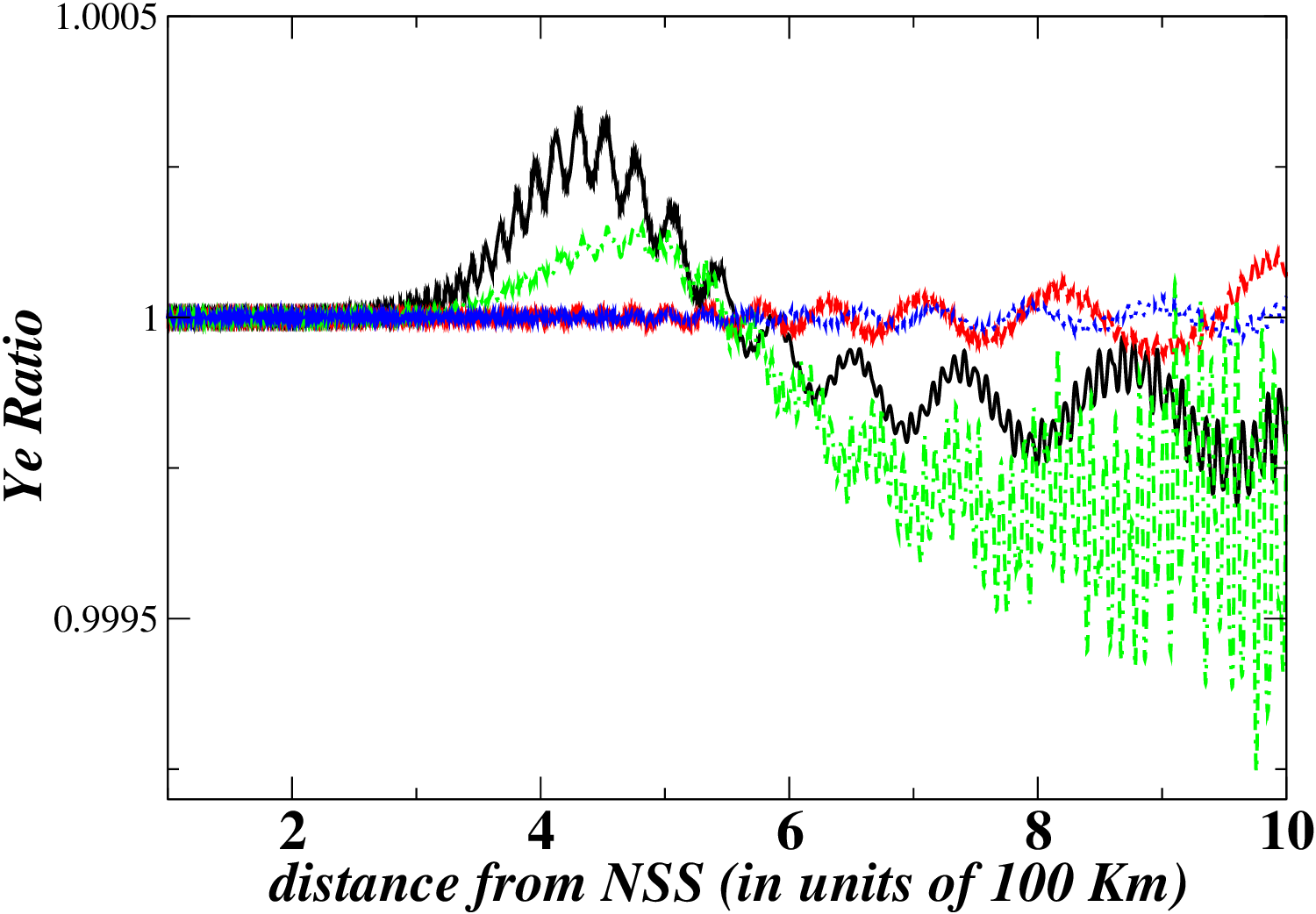}
  \caption{Ratio of the electron fraction as a function of the distance in the supernova for a non-zero over a zero delta phase. The different curves correspond to normal hierarchy and large (solid),
or small (dashed) $\theta_{13}$, and to inverted and large (dot-dashed) or small (dotted) $\theta_{13}$
\cite{Balantekin:2007es}. \label{fig:Ye}}
\vspace{.2cm}
\end{figure}

\begin{figure}
  \includegraphics[height=.25\textheight]{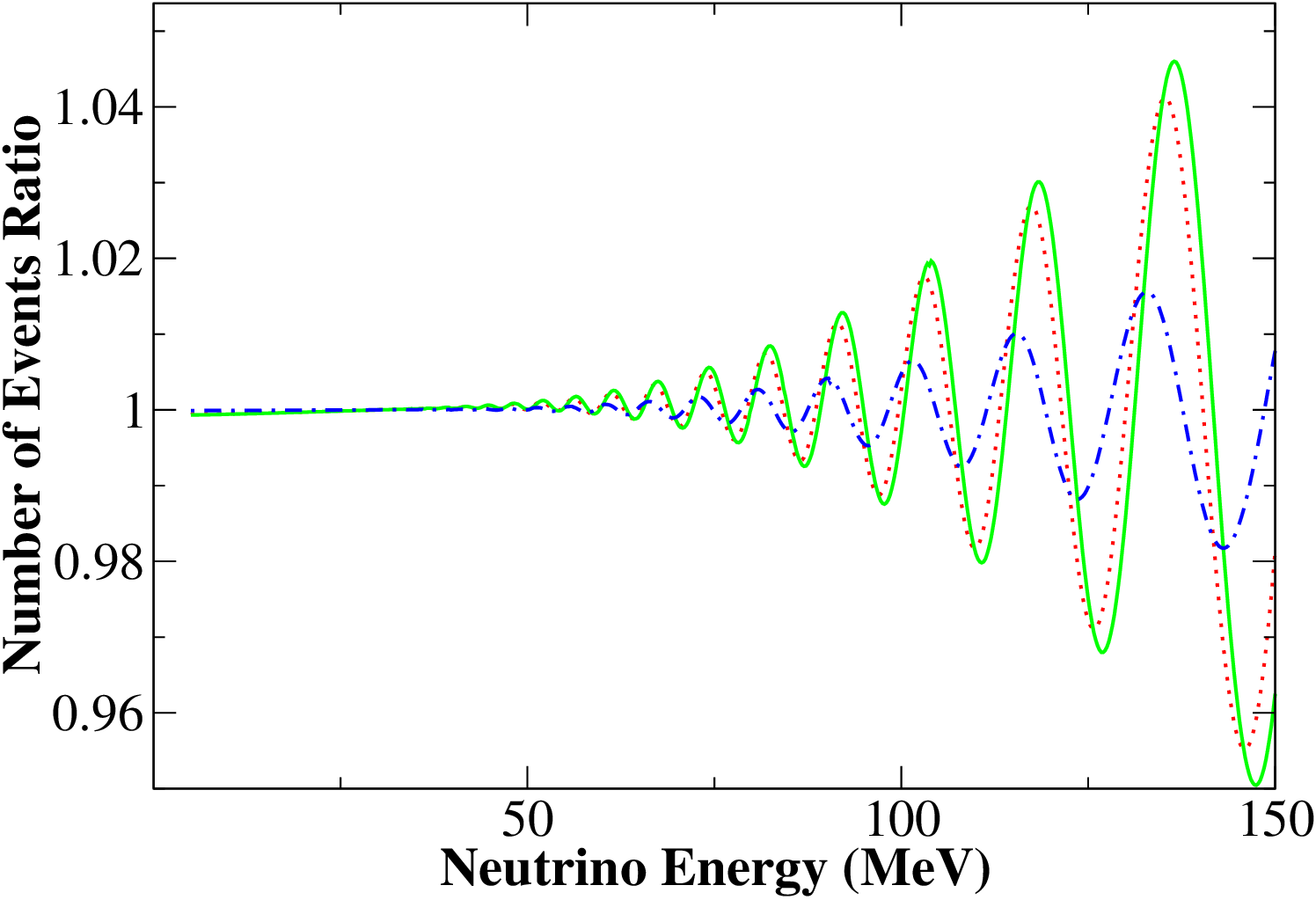}
  \caption{Ratio of the number of events, for a non-zero over a zero $\delta$ phase, associated to electron anti-neutrino scattering on protons in a  
water \v{C}erenkov detector on Earth, such as Super-Kamiokande (22.5 kton fiducial volume). 
A 100 $\%$ efficiency is assumed. The curves show the effects induced by a non-zero CP violating phase, i.e. $\delta=180^{\circ}$ (solid), $135^{\circ}$(dashed), $45^{\circ}$ (dot-dashed)  \cite{Balantekin:2007es}. \label{fig:events}}
\vspace{.2cm}
\end{figure}

We have also investigated indirect impact on the r-process by calculating how the electron fraction is modified: the effects turn out to be tiny (Figure \ref{fig:Ye}). Finally we have determined the possible impact on the neutrino signal in a detector on Earth\footnote{To get this result we calculate the neutrino fluxes far our the supernova
and average them when vacuum is reached.}. 
While the $\delta$ effect on the total number of events is of the order 
of 2$\times 10^{-4}$, variations
up to 5$\%$ are obtained for the events as a function of neutrino energy (Figure \ref{fig:events}).
\footnote{This conclusion is at variance with section
3.4 of Ref.\cite{Akhmedov:2002zj} where the authors find that are no CP violating
effects in a detector on Earth, if muon and tau neutrinos have equal or different
luminosities at the neutrinosphere. Note that they use the matter eigenstates (in their derivation) which is equal to the physical flavour basis used in our calculations only at very high densities.}

\section{Electron neutrino detection and future neutrino-nucleus measurements}
Neutrino-nucleus interactions is a topic of current great interest for e.g. understanding
the neutrino detector's response, the nucleosynthesis of heavy elements and of the isospin and
spin-isospin nuclear response (see e.g. the reviews \cite{Balantekin:2003ip,Volpe:2004dg}). For example it is shown in \cite{Volpe:2005iy}
that such measurements furnish a new constraints for neutrinoless
double-beta decay predictions that are at present affected by significant variations. 
In supernova neutrino observatories various nuclei are exploited as neutrino targets: carbon in scintillator detectors, oxygen in water \v{C}erenkov, lead in lead-based detectors,
argon in liquid argon, deuteron in heavy water. Deuteron is by far the best case since it can
be predicted theoretically with errors of a few percent in Effective Field Theory.
Carbon is the best studied case experimentally although the inclusive cross sections are still affected by
significant variations (see e.g. \cite{Volpe:2000zn}). The only other available measurement is on iron.
Theoretical predictions employing
 microscopic models (in particular
the shell model and the QRPA) can present discrepancies arising for example from the treatment of the continuum and from
different contributions of the forbidden multipoles (the spin-dipole and higher multipoles). 

Future measurements with very intense low energy neutrino beams can provide important
information that would put the theoretical predictions on neutrino-nucleus
cross sections on a firmer ground. Two avenues are
possible: either an intense conventional source (neutrinos from muon decay-at-rest) 
or low energy beta-beams\footnote{Beta-beams are pure and intense neutrino beams of well known fluxes that exploit the beta-decay of boosted radioactive ions \cite{Zucchelli:2002sa}.} 
\cite{Volpe:2003fi}. Intense conventional sources will be available 
at $\nu$ SNS at Los Alamos \cite{sns} and at JPARC. A low energy beta-beam facility might be
established, in particular if a beta-beam for the search CP violation in the lepton sector is built. 
It is shown in \cite{Serreau:2004kx} that this might require a devoted storage
ring while a less expensive option can be to use one/two detectors at off-axis of the storage
ring planned for CP violation \cite{Lazauskas:2007va}. By varying the Lorentz boost of the ions one can try to disentangle
the contribution of the forbidden multipoles for which little information is available right now \cite{Volpe:2003fi,McLaughlin:2004va,Lazauskas:2007bs}. Besides improving our knowledge of
the supernova detector's response, other interesting supernova neutrino applications are
possible wth low energy beta-beams. 
In particular 
combining measurements at different boosts one could reconstruct the neutrino signal 
from a supernova explosion  \cite{Jachowicz:2006xx,Jachowicz:2008kx}.

\section{Conclusions}
Supernova neutrino experiments can bring crucial information, if a galactic
supernova explosion occurs, or a  future  large-scale observatory observe an extragalactic event and measure relic supernova neutrinos. Such observations can help
unravelling important features of the supernova dynamics, or of the still unknown neutrino properties. Important theoretical developments are also ongoing, in particular in our understanding
of neutrino propagation in dense matter, since the neutrino-neutrino interaction included only recently in the calculations
produces completely new features in some cases.

Here we have discussed the possible indirect and direct (on a neutrino
signal) impact of CP violation in the lepton sector in the supernova
environment. 
Our numerical calculations show that effects on the electron (anti-)neutrino fluxes  
within the supernova can be as large as factors 2-4, if  
muon and tau neutrinos have different luminosities at the neutrinosphere (due e.g. to physics beyond the Standard Model
such as flavour changing interactions). Possible effects on the electron fraction, relevant for the r-process, are tiny; while
variations up to 5$\%$ on the events as a function of neutrino energy are found in a detector on Earth.  

One of the difficulties of this domain is that the information on the supernova dynamics, on the neutrino properties and
the star formation rate (for relic neutrinos) is all entangled. It is therefore important to have observatories sensitive both to electron neutrinos and anti-neutrinos, to extract as much information as possible from observations. In this respect, important progress still need to be made on our knowledge of the neutrino-nucleus interaction, necessary
for the electron neutrino detection through charged-current reactions.


\begin{theacknowledgments}
The authors acknowledge the support from the ANR-05-JCJC-0023 "Non standard
neutrino  properties and their impact in astrophysics and cosmology".
\end{theacknowledgments}

\end{document}